\documentclass[aps,prd,a4paper,twocolumn,amsmath,showpacs,superscriptaddress,nofootinbib,preprintnumbers]{revtex4-1}
\pdfoutput=1
\usepackage{amssymb,amsmath,latexsym,mathrsfs}
\usepackage[sort&compress]{natbib}
\usepackage{graphicx,subfigure}
\usepackage{epsfig}
\usepackage{varioref,xr-hyper}
\usepackage{color}
\usepackage{multirow}
\usepackage{array}
\usepackage{hyperref}
\usepackage{wasysym}
\usepackage{color}
\usepackage{float}
\usepackage{xcolor}
\usepackage[utf8]{inputenc}
\usepackage[T1]{fontenc}

\begin{document}


\title{Reconciling $H_0$ tension in a six parameter space?}


\author{Supriya Pan}
\email{supriya.maths@presiuniv.ac.in}
\affiliation{Department of Mathematics, Presidency University, 86/1 College Street, Kolkata 700073, India}

\author{Weiqiang Yang}
\email{d11102004@163.com}
\affiliation{Department of Physics, Liaoning Normal University, Dalian, 116029, People's Republic of China}

\author{Eleonora Di Valentino}
\email{eleonora.divalentino@manchester.ac.uk}
\affiliation{Jodrell Bank Center for Astrophysics, School of Physics and Astronomy,  University of  Manchester, Oxford Road, Manchester, M13 9PL, United Kingdom}

\author{Arman Shafieloo}
\email{shafieloo@kasi.re.kr}
\affiliation{Korea Astronomy and Space Science Institute, Daejeon 34055, Korea}
\affiliation{University of Science and Technology, Yuseong-gu 217 Gajeong-ro, Daejeon 34113, Korea} 

\author{Subenoy Chakraborty}
\email{schakraborty.math@gmail.com}
\affiliation{Department of Mathematics, Jadavpur University, Kolkata 700032, West Bengal,  India}


\begin{abstract}
Consistent observations indicate that some of the important cosmological parameters measured through the local observations are in huge tension with their measurements from the global observations (within the minimal $\Lambda$CDM cosmology). The tensions in those cosmological parameters have been found to be either weakened or reconciled with the introduction of new degrees of freedom that effectively increases the underlying parameter space compared to the minimal $\Lambda$CDM cosmology. It might be interesting to investigate the above tensions within the context of an emergent dark energy scenario proposed recently by Li and Shafieloo~\cite{Li:2019yem}. We find that the tension on $H_0$ is clearly alleviated within 68\% confidence level with an improvement of the $\chi^2$ for CMB, for the above emergent dark energy model having only six free parameters similar to the spatially flat $\Lambda$CDM model.  The tension on $H_0$ is still alleviated for every combined datasets considered in the work, however, such alleviation occurs by worsening the $\chi^2$ compared to the $\chi^2$ for $\Lambda$CDM model obtained for the same combined dataset. 
\end{abstract}

\pacs{98.80.-k, 95.36.+x, 95.35.+d, 98.80.Es}
\maketitle
\section{Introduction}

According to a series of distinct observational data,  such as cosmic microwave background (CMB) radiation~\cite{Ade:2015xua,Aghanim:2018eyx} and baryon acoustic oscillations (BAO) distance  measurements~\cite{Beutler:2011hx,Ross:2014qpa,Alam:2016hwk}, the $\Lambda$-cold-dark-matter ($\Lambda$CDM) cosmology is one of the best cosmological descriptions for the currently accelerated expansion of the universe, but on the other hand, it has been diagnosed with a number of severe problems. Apart from its inherent cosmological constant problem, the estimations of some important cosmological parameters in $\Lambda$CDM based cosmological framework exhibit tensions with respect to their estimations by other measurements. For instance, the estimation of the Hubble constant $H_0$ from $\Lambda$CDM based Planck's mission \cite{Aghanim:2018eyx} is more than $4\sigma$ apart from its estimation by the SH0ES collaboration~\cite{Riess:2019cxk}, more than $5\sigma$ if combined with the H0liCOW collaboration result~\cite{Wong:2019kwg}, and around $4.5\sigma$ for considering the cosmographic expansion of the luminosity distance~\cite{Camarena:2019moy}.  In general, there is an $H_0$ tension between late time and early time estimations, ranging from $4.5\sigma$ to $6.3\sigma$~\cite{Riess:2020sih}. On the other hand, also the estimation of the $S_8~(\equiv \sigma_8 \sqrt{\Omega_{m0}/0.3})$ parameter from Planck in a $\Lambda$CDM scenario \cite{Ade:2015xua} is in tension {\bf at about 2.5$\sigma$} with the cosmic shear measurements by different missions, for instance, KiDS-450~\cite{Kuijken:2015vca,Hildebrandt:2016iqg,Conti:2016gav}, DES-Y1~\cite{Abbott:2017wau,Troxel:2017xyo} and CFHTLenS~\cite{Heymans:2012gg, Erben:2012zw,Joudaki:2016mvz},  or Lyman-$\alpha$ data \cite{Palanque-Delabrouille:2019iyz}, and about $3.2\sigma$ tension with the combination of KiDS+VIKING-450 and DES-Y1~\cite{Asgari:2019fkq}.
However, it should be noted that there are many other measurements too in agreement with Planck about $H_0$ and $S_8$, like for example the BAO or the Tip of the Red Giant Branch estimates of $H_0$~\cite{Freedman:2020dne}, or the HSC collaboration value of $S_8$~\cite{Hamana:2019etx}, and that these two tensions do not have the same level of statistical significance.

Whether such tensions call for a new physics  \cite{Mortsell:2018mfj,Vagnozzi:2019ezj} or they are arising due to the systematics \cite{Efstathiou:2013via} are not clearly understood at this stage. However, undoubtedly, the $H_0$ and $\sigma_8$ tensions are two primary issues for modern cosmology and should be carefully investigated.

Since $\Lambda$CDM is unable to explain these issues \footnote{Obviously, although not very probable, these problems in $\Lambda$CDM we are worried about might be due to undetected systematics in some of the experiments. }, an usual approach is to consider the cosmological models beyond $\Lambda$CDM. Following this motivation, several extensions of the $\Lambda$CDM cosmology have been introduced with a  possible solution to the $H_0$ tension~\cite{DiValentino:2015ola,DiValentino:2016hlg,Kumar:2017dnp,DiValentino:2017iww,DiValentino:2017zyq,Renk:2017rzu,DiValentino:2017gzb,DiValentino:2017oaw,Fernandez-Arenas:2017isq,DiValentino:2017rcr,Khosravi:2017hfi,Sola:2017znb,Nunes:2018xbm,Yang:2018euj,Colgain:2018wgk,DEramo:2018vss,Yang:2018uae,Guo:2018ans,Yang:2018qmz,Poulin:2018cxd,Banihashemi:2018oxo,Banihashemi:2018has,Zhang:2018air,Kreisch:2019yzn,Martinelli:2019dau,Vattis:2019efj,Kumar:2019wfs,Agrawal:2019lmo,Yang:2019jwn,Yang:2019qza,Yang:2019uzo,DiValentino:2019exe,Desmond:2019ygn,Yang:2019nhz,Pan:2019gop,Visinelli:2019qqu,Martinelli:2019krf,Cai:2019bdh,Schoneberg:2019wmt,Shafieloo:2016bpk,Li:2019san} and $\sigma_8$ tension as well  \cite{Pourtsidou:2016ico,An:2017crg,Gomez-Valent:2017idt,DiValentino:2018gcu,Kumar:2019wfs,Gomez-Valent:2018nib,Kazantzidis:2018rnb,Hazra:2018opk,Kazantzidis:2019dvk} (also see \cite{Macaulay:2013swa} where the authors reported lower $\sigma_8$ compared to Planck). However, extended cosmological models naturally include extra free parameters compared to the six parameter $\Lambda$CDM scenario, and are therefore disfavoured with respect to it. It has thus been a natural search for some alternative cosmological model having same number of free parameters as in $\Lambda$CDM but having the ability to solve or reconcile the tension on of the two important parameters, namely, $H_0$ and $\sigma_8$.

In the present article we work with a dynamical emergent dark energy model, recently introduced  in \cite{Li:2019yem}, that has exactly same number of free parameters as in $\Lambda$CDM model. We investigate the model considering its evolution at the level of background and perturbations and constrain it using the presently available cosmological datasets including Planck 2015 cosmic microwave background (CMB) radiation, Pantheon sample of the Supernovae Type Ia, Baryon acoustic oscillations distance measurements, and the recently released local estimation of the Hubble constant by Riess et al. \cite{Riess:2019cxk}. Our analyses clearly show that the tension on $H_0$ is reconciled within 68\% confidence-level for this model \cite{Li:2019yem}.  This is one of the  key results of this paper because so far we are aware of the literature, probably this is the first time we are reporting the reconciliation of $H_0$ tension in a six parameter space, improving the $\chi^2$ for CMB.

The work has been organized in the following way. In section \ref{sec-2} we briefly   
discuss the basic governing equations for the introduced dynamical dark energy model in a spatially flat Friedmann-Lema\^{i}tre-Robertson-Walker (FLRW) universe. In section \ref{sec-data} we present the observational data and the methodology for this paper. After that in section \ref{sec-results} we discuss the main results extracted from this model. Finally, we close the work in section \ref{sec-discuss} with  a short summary of entire results.

\section{Phenomenologically Emergent Dark Energy}
\label{sec-2}

We consider a spatially flat Friedmann-Lema\^{i}tre-Robertson-Walker (FLRW) metric to describe the geometrical configuration of the universe. We also consider that the gravitational sector of the universe is well described by the Einstein gravity where matter is minimally coupled to it. Additionally, we further assume that none of the fluids are interacting with each other, at least non-gravitationally. 
So, if the content of the universe is comprised of radiation, pressureless matter sector (baryons+cold dark matter) and a dark energy fluid \footnote{Let us note that here we fix the total neutrino mass to $M_{\nu}=0.06\,{\rm eV}$ according to the Planck mission. This is certainly justified through the tight upper limits available on $M_{\nu}$~\cite{Palanque-Delabrouille:2015pga,Giusarma:2016phn,Vagnozzi:2017ovm,Giusarma:2018jei,Aghanim:2018eyx}. }, then in the background of a spatially flat FLRW universe, one can write down the Hubble equation as 

\begin{eqnarray}
H^2 = H_0^2  \Bigl[\Omega_{r0} (1+z)^4 + \Omega_{m0} (1+z)^3 + \Omega_{DE} (z)\Bigr]
\end{eqnarray}    
where $H$ is the Hubble parameter of the FLRW universe, $\Omega_{r0}$ is the density parameter for radiation, $\Omega_{m0}$ is the density parameter for matter (baryons+cold dark matter) and $\Omega_{DE} (z)$ is the dark energy density parameter. The dark energy density parameter can be solved as 

\begin{eqnarray}
\Omega_{DE} (z) = \Omega_{DE,0} \; \exp \left[ 3 \; \int_{0}^{z} \frac{1+w_{DE}(z^\prime)}{1+z^\prime} dz^\prime \right]
\end{eqnarray}
where $\Omega_{DE,0}$ is the current value of $\Omega_{DE}$; $w_{DE} (z) = p_{DE} (z)/\rho_{DE} (z)$, is the equation-of-state of the dark energy fluid. There are various ways to depict the evolution of the universe $-$ either by prescribing the equation-of-state of the dark energy, or by providing the density parameter for dark energy. 

In this work we shall consider the second approach  recently proposed in \cite{Li:2019yem}: 

\begin{eqnarray}\label{model}
\Omega_{DE} (z) = \Omega_{DE,0} \Bigl[1- \tanh (\log_{10} (1+z)) \Bigr]
\end{eqnarray}
where $\Omega_{DE,0} =  1- \Omega_{m0}-\Omega_{r0}$ and $1+z  = a_0 a^{-1} = a^{-1}$ (without any loss of generality we set $a_0$, the current value of the scale factor to be unity, i.e., $a_0 = 1$). 
As already argued in \cite{Li:2019yem} this model is similar to the $\Lambda$CDM one in the sense that both the models have six free parameters. So, from the statistical ground the models are same. Certainly, it will be interesting to investigate such phenomenological model having same number of free parameters in light of the latest observations. 

We should mention here that although the model proposed in ~\cite{Li:2019yem} might seems to be ad hoc, it has interesting phenomenological properties that dark energy acts as an emergent phenomena having a symmetrical behavior with respect to the logarithm of the scale factor. This model has already gained attention from the community as a competitor of other well known cosmological models \cite{Arendse:2019hev} and might provide valuable hints for theoretical interpretations. While in Ref.~\cite{Li:2019yem}, the authors present its observational constraints at the level of background, its evolution at the level of perturbations is worth to understand. 
In the current work we therefore aim to extend this study by analysing its behaviour at the  level of perturbations and the cosmological tensions which have already been a serious issue in the context of $\Lambda$CDM cosmology.

 Since there is no interaction between any two fluids under consideration, hence, using the conservation equation for dark energy, namely, 
$$\dot{\rho}_{DE} (z) + 3 H (1+w_{DE} (z))\rho_{DE} (z) = 0,$$ 
one can derive the dark energy equation-of-state as 
\begin{eqnarray}\label{de-eos}
w_{DE} (z) =  -1 + \frac{1}{1+z} \times \frac{d\; \ln \Omega_{DE} (z)}{dz},
\end{eqnarray}
which for the present model in (\ref{model}) takes the form \cite{Li:2019yem}
\begin{align}\label{eos}
w_{DE} (z) = -1 -\frac{1}{3 \ln 10} \times \Bigl[ 1+ \tanh \bigl(\log_{10} (1+z)\bigr) \Bigr]
\end{align}
Thus, for any prescribed $\Omega_{DE} (z)$ one can derive the dark energy equation-of-state.  As explained in \cite{Li:2019yem}, the equation-of-state (\ref{de-eos}) has an interesting symmetrical feature.  
From (\ref{eos}) one can see that at early time, i.e. for $z \rightarrow \infty$, $w_{DE} \rightarrow -1 -\frac{2}{3 \ln 10}$ and for $z \rightarrow -1$ (far future), $w_{DE} \rightarrow -1$. And at preset time (i.e. $z = 0$), we see $w_{DE} = -1 -\frac{1}{3 \ln 10}$, that means a phantom dark energy equation of state. Note that as described briefly in \cite{Li:2019yem}, the pivot point of transition in this model can be considered to be the redshift of matter-dark energy densities equality.   
For the present model in (\ref{model}), dark energy has no effective presence in the past as shown in Fig. 1  of \cite{Li:2019yem}, while it emerges at present time, therefore, by the authors of \cite{Li:2019yem}, this model has been named as Phenomenologically Emergent Dark Energy (PEDE) model and we use the same name throughout this article.  So, having presented the equations above, the background evolution of the PEDE model is clearly understood. Concerning the perturbations equations for this model, our treatment is does not involve anything new because as the equation of state of DE under this PEDE model is less than $-1$ and the cosmological scenario does not involve any non-gravitational interaction between any two fluids, hence, the perturbations equations will be exactly similar to the equations for a non-interacting phantom DE equation of state. We refer to the work \cite{Ma:1995ey} for implementing the perturbations equations,  specially eqn. (29) [for synchronous gauge] or eqn. (30) [for conformal Newtonian gauge].

\section{Observational data}
\label{sec-data}

In this section we describe the main observational data that are used to constrain the proposed dark energy model. 

\begin{enumerate}

\item \textbf{Cosmic Microwave Background (CMB)}: The Cosmic Microwave Background measurements are one of the potential data to unveil the nature of the dark universe. Here we make use of the Planck 2015 data~\cite{Adam:2015rua, Aghanim:2015xee} that include both high-$\ell$ ($30\le \ell \le 2508$) TT and low-$\ell$ ($2\le \ell \le 29$) TT likelihoods. We also consider the Planck polarization likelihood in the low-$\ell$ multipole regime ($2\le \ell \le 29$) as well as the high-multipole ($30\le \ell \le 1996$) EE and TE likelihoods.

\item \textbf{Baryon acoustic oscillation (BAO) distance measurements}:  We use 6dFGS~\cite{Beutler:2011hx}, SDSS-MGS~\cite{Ross:2014qpa}, and 
BOSS DR12~\cite{Alam:2016hwk} surveys, as 
considered by the Planck collaboration~\cite{Aghanim:2018eyx}.

\item \textbf{Supernovae Type Ia (Pantheon)}:  The Supernovae Type Ia (SNIa) were the first standard candles that signaled for an accelerating universe. In this work we make use of the Pantheon sample, the latest compilation of SNIa, comprising 1048 data points in the redshift region $z \in [0.01, 2.3]$ ~\cite{Scolnic:2017caz}. 

\item \textbf{Hubble constant (R19)}: We include the recent estimation of the Hubble constant, $H_0 = 74.03 \pm 1.42$ km/s/Mpc at $68\%$ CL~\cite{Riess:2019cxk}, which is in tension ($4.4 \sigma$) with CMB estimation within the minimal cosmological model $\Lambda$CDM. 

\item \textbf{Dark energy survey (DES):} We consider the  $3\times2$pt analysis of the first-year of the Dark Energy Survey measurements~\cite{Troxel:2017xyo, Abbott:2017wau, Krause:2017ekm}, as adopted by the Planck collaboration in~\cite{Aghanim:2018eyx}.

\item \textbf{Lensing:} We use the CMB lensing reconstruction power spectrum obtained from the CMB trispectrum analysis~\cite{Ade:2015zua}.

\end{enumerate}

To perform the numerical analysis we use the markov Chain Monte Carlo (MCMC) package \texttt{CosmoMC}~\cite{Lewis:2002ah,Lewis:1999bs} which is equipped with a convergence statistics by Gelman and Rubin. This \texttt{CosmoMC} package includes the support for Planck 2015 likelihood~\cite{Aghanim:2015xee}. The parameter space that we will consider has six parameters similarly to the $\Lambda$CDM model. In particular we have the following parameter space 
\begin{align}
\mathcal{P} \equiv\Bigl\{\Omega_{b}h^2, \Omega_{c}h^2, 100\theta_{MC}, \tau, n_{s}, log[10^{10}A_{s}]\Bigr\}~,
\label{eq:parameter_space}
\end{align}
where $\Omega_{b} h^2$ is the physical density for baryons, $\Omega_{c}h^2$ is the physical density for CDM, $\theta_{MC}$ is the ratio of sound horizon to the angular diameter distance, $\tau$ denotes the reionization optical depth, $n_{s}$ is the scalar spectral index, and $A_S$ is the amplitude of the primordial scalar power spectrum. In  Table \ref{tab:priors} we display the priors that are imposed on various free parameters during the statistical analysis.

\begin{table}
\begin{center}
\renewcommand{\arraystretch}{1.4}
\begin{tabular}{|c@{\hspace{1 cm}}|@{\hspace{1 cm}} c|}
\hline
\textbf{Parameter}                    & \textbf{Prior}\\
\hline\hline
$\Omega_{b} h^2$             & $[0.005,0.1]$\\
$\Omega_{c} h^2$             & $[0.01,0.99]$\\
$\tau$                       & $[0.01,0.8]$\\
$n_s$                        & $[0.5, 1.5]$\\
$\log[10^{10}A_{s}]$         & $[2.4,4]$\\
$100\theta_{MC}$             & $[0.5,10]$\\ 
\hline
\end{tabular}
\end{center}
\caption{Priors imposed on various free parameters of the PEDE and $\Lambda$CDM cosmological models. Recall that this model has same number of parameters as in flat $\Lambda$CDM model.}
\label{tab:priors}
\end{table}
\begingroup                                                                                                                     
\squeezetable                                                                                                                   
\begin{center}                                                                                                                  
\begin{table*}
\scalebox{0.9}{                                              
\begin{tabular}{ccccccccccccc}                                                                                                            
\hline\hline                                                                                                                    
Parameters & CMB & CMB+BAO & CMB+Pantheon & CMB+R19 & CMB+DES & CMB+Lensing \\ \hline

$\Omega_c h^2$ & $    0.1191_{-    0.0014-    0.0027}^{+    0.0014+    0.0027}$ & $    0.1210_{-    0.00099-    0.0020}^{+    0.0010+    0.0019}$ & $    0.1213_{-    0.0013-    0.0024}^{+    0.0013+    0.0025}$ & $    0.1186_{-    0.0012-    0.0025}^{+    0.0013+    0.0025}$  & $    0.1173_{-    0.0012-    0.0023}^{+    0.0012+    0.0023}$ & $    0.1186_{-    0.0013-    0.0027}^{+    0.0013+    0.0026}$  \\

$\Omega_b h^2$ & $    0.02227_{-    0.00016-    0.00030}^{+    0.00016+    0.00031}$ & 
$    0.02213_{-    0.00013-    0.00026}^{+    0.00013+    0.00027}$ & $    0.02211_{-    0.00014-    0.00028}^{+    0.00014+    0.00029}$ &  $    0.02231_{-    0.00016-    0.00029}^{+    0.00015+    0.00029}$ & $    0.02240_{-    0.00015-    0.00029}^{+    0.00015+    0.00030}$  & $    0.02228_{-    0.00016-    0.00031}^{+    0.00016+    0.00030}$  \\

$100\theta_{MC}$ & $    1.04077_{-    0.00032-    0.00063}^{+    0.00032+    0.00064}$ & $    1.04056_{-    0.00029-    0.00058}^{+    0.00030+    0.00059}$ & $    1.04053_{-    0.00031-    0.00063}^{+    0.00032+    0.00063}$ & $    1.04086_{-    0.00033-    0.00058}^{+    0.00030+    0.00063}$  & $    1.04098_{-    0.00034-    0.00062}^{+    0.00031+    0.00064}$ & $    1.04087_{-    0.00031-    0.00067}^{+    0.00033+    0.00063}$  \\

$\tau$ & $    0.079_{-    0.017-    0.034}^{+    0.017+    0.033}$ & $    0.070_{-    0.016-    0.031}^{+    0.016+    0.032}$ & $    0.069_{-    0.017-    0.033}^{+    0.017+    0.032}$  & $    0.082_{-    0.017-    0.035}^{+    0.017+    0.033}$ &  $    0.074_{-    0.017-    0.033}^{+    0.017+    0.033}$  & $    0.060_{-    0.013-    0.027}^{+    0.013+    0.027}$   \\

$n_s$ & $    0.9664_{-    0.0046-    0.0088}^{+    0.0046+    0.0089}$ & $    0.9616_{-    0.0038-    0.0076}^{+    0.0038+    0.0076}$ & $    0.9608_{-    0.0042-    0.0083}^{+    0.0042+    0.0083}$  & $    0.9678_{-    0.0043-    0.0086}^{+    0.0044+    0.0086}$ & $    0.9702_{-    0.0043-    0.0081}^{+    0.0044+    0.0084}$   & $    0.9671_{-    0.0044-    0.0090}^{+    0.0045+    0.0090}$  \\

${\rm{ln}}(10^{10} A_s)$ & $    3.090_{-    0.033-    0.067}^{+    0.034+    0.064}$ & 
$    3.078_{-    0.032-    0.063}^{+    0.032+    0.062}$ & $    3.075_{-    0.032-    0.065}^{+    0.033+    0.062}$ & $    3.095_{-    0.034-    0.070}^{+    0.035+    0.064}$ &  $    3.077_{-    0.033-    0.065}^{+    0.032+    0.064}$  & $    3.052_{-    0.024-    0.050}^{+    0.024+    0.051}$  \\

$\Omega_{m0}$ & $    0.270_{-    0.0082-    0.016}^{+    0.0083+    0.017}$ &  $    0.281_{-    0.0060-    0.012}^{+    0.0061+    0.012}$  & $    0.283_{-    0.0079-    0.015}^{+    0.0079+    0.015}$ & $    0.266_{-    0.0073-    0.014}^{+    0.0073+    0.015}$ &  $    0.259_{-    0.0068-    0.013}^{+    0.0069+    0.013}$ & $    0.267_{-    0.0077-    0.016}^{+    0.0078+    0.015}$  \\

$\sigma_8$ & $    0.876_{-    0.014-    0.028}^{+    0.014+    0.028}$ & $    0.875_{-    0.014-    0.027}^{+    0.014+    0.027}$ & $    0.874_{-    0.014-    0.027}^{+    0.014+    0.027}$ &  $    0.877_{-    0.014-    0.030}^{+    0.015+    0.027}$  & $    0.865_{-    0.013-    0.026}^{+    0.013+    0.026}$ & $    0.858_{-    0.0090-    0.019}^{+    0.0090+    0.019}$   \\

$H_0$ & $   72.58_{-    0.80-    1.5}^{+    0.79+    1.6}$ & $   71.55_{-    0.57-    1.1}^{+    0.55+    1.1}$  & $   71.36_{-    0.71-    1.4}^{+    0.71+    1.4}$ & $   72.92_{-    0.71-    1.4}^{+    0.72+    1.4}$ & $   73.65_{-    0.71-    1.3}^{+    0.69+    1.4}$ & $   72.86_{-    0.75-    1.5}^{+    0.76+    1.6}$  \\

$S_8$ & $0.831^{+0.017+0.035}_{-0.018-0.035}$ & $0.846^{+ 0.016+ 0.031}_{- 0.015- 0.031}$ & $0.849^{+0.016+0.033}_{-0.017-0.033}$ & $0.826^{+0.016+0.033}_{-0.017-0.034}$ & $0.803^{+0.013+0.026}_{-0.013-0.026}$ & $0.809^{+0.012+0.024}_{-0.012-0.024}$ \\

\hline 

$\chi^2$ & 12962.468 & 12976.566 & 14010.266 & 12963.448 & 13485.282 & 12974.282\\

\hline\hline                                                                                                                    
\end{tabular}
}                                                                                                                   
\caption{We report the 68\% and 95\% CL constraints on the free and derived parameters of the cosmic scenario driven by the Phenomenologically Emergent Dark Energy model (\ref{model}) using various observational datasets. In the last row of the table we also display the best-fit values of $\chi^2$ for all the observational datasets. }
\label{tab:PEDE}                                                                                                   
\end{table*}                                                    
\end{center}
\endgroup
\begin{figure*}
\includegraphics[width=0.9\textwidth]{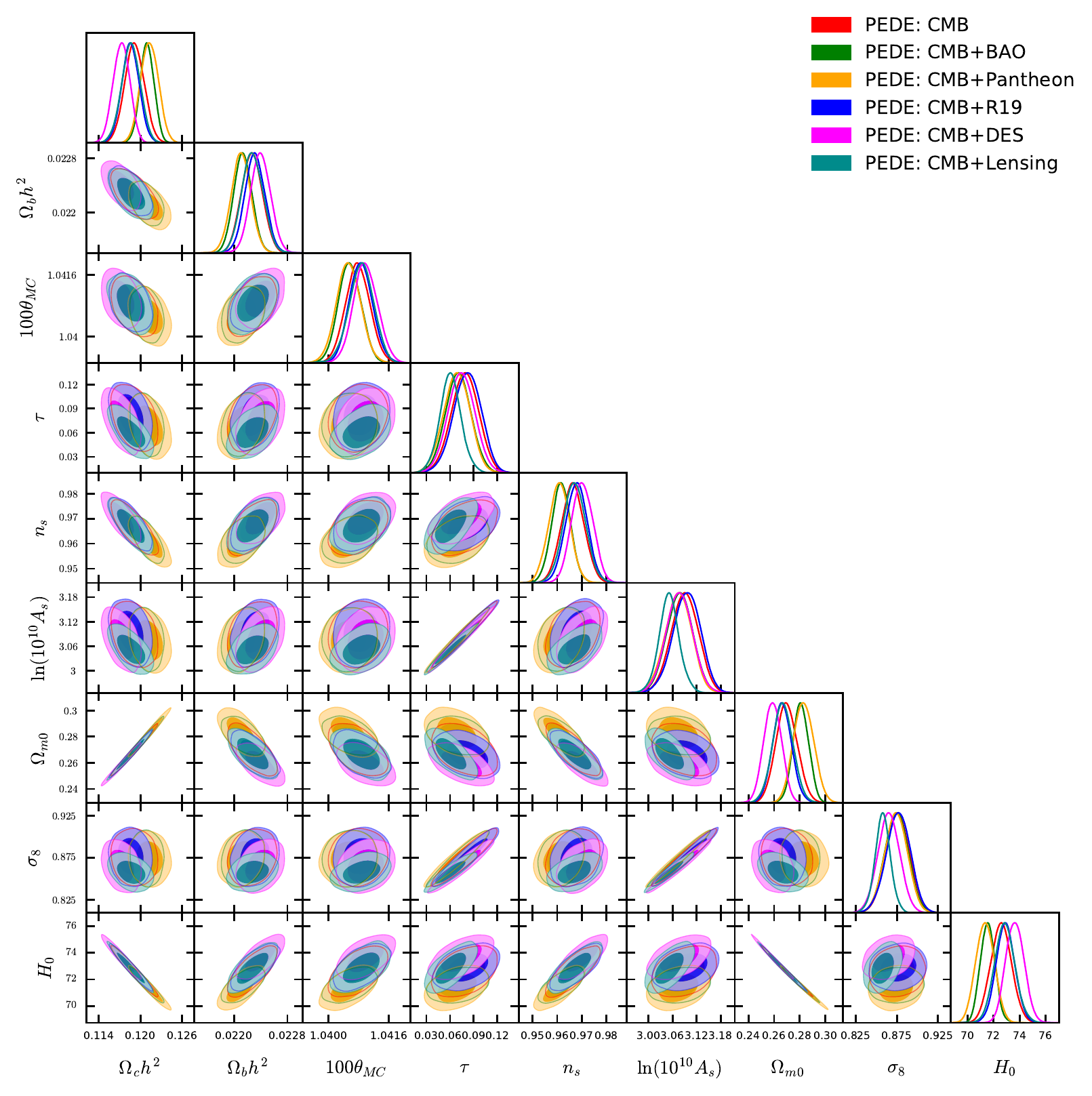}
\caption{1D marginalized posterior distributions of all the free and derived parameters of the PEDE model as well as 2D joint contours at 68\% and 95\% confidence level are shown for various observational datasets.  }
\label{fig:pede}
\end{figure*}
\begingroup                                                 
\squeezetable                                                   
\begin{center}                                                  
\begin{table*}
\scalebox{0.9}{                                                 
\begin{tabular}{ccccccccccccccc}                                  
\hline\hline                                                                                                                    
Parameters & CMB & CMB+BAO & CMB+Pantheon & CMB+R19 & CMB+DES & CMB+Lensing \\ \hline

$\Omega_c h^2$ & $    0.1192_{-    0.0015-    0.0027}^{+    0.0014+    0.0028}$ & $    0.1187_{-    0.0010-    0.0020}^{+    0.0010+    0.0020}$ & $    0.1190_{-    0.0014-    0.0025}^{+    0.0013+    0.0025}$ & $    0.1170_{-    0.0013-    0.0025}^{+    0.0013+    0.0025}$ & $    0.1170_{-    0.0012-    0.0024}^{+    0.0012+    0.0024}$ & $    0.1188_{-    0.0013-    0.0027}^{+    0.0013+    0.0026}$  \\

$\Omega_b h^2$ & $    0.02225_{-    0.00016-    0.00030}^{+    0.00016+    0.00030}$ & $    0.02229_{-    0.00014-    0.00027}^{+    0.00014+    0.00026}$  & $    0.02227_{-    0.00015-    0.00030}^{+    0.00015+    0.00029}$ & $    0.02245_{-    0.00015-    0.00030}^{+    0.00015+    0.00031}$  & $    0.02241_{-    0.00014-    0.00028}^{+    0.00014+    0.00029}$ & $    0.02225_{-    0.00015-    0.00029}^{+    0.00015+    0.00030}$  \\

$100\theta_{MC}$ & $    1.04076_{-    0.00033-    0.00062}^{+    0.00033+    0.00063}$ & $    1.04085_{-    0.00030-    0.00060}^{+    0.00030+    0.00059}$ & $    1.04079_{-    0.00032-    0.00063}^{+    0.00032+    0.00064}$ & $    1.04108_{- 0.00031-    0.000636}^{+    0.00031+    0.00064}$  & $    1.04103_{-    0.00031-    0.00059}^{+    0.00030+    0.00062}$  & $    1.04083_{-    0.00032-    0.00063}^{+    0.00033+    0.00065}$  \\

$\tau$ & $    0.080_{-    0.017-    0.034}^{+    0.017+    0.033}$ & $    0.083_{-    0.017-    0.032}^{+    0.017+    0.032}$  & $    0.082_{-    0.017-    0.033}^{+    0.017+    0.033}$  & $    0.092_{-    0.017-    0.033}^{+    0.017+    0.032}$ &  $    0.075_{-    0.017-    0.034}^{+    0.018+    0.034}$  & $    0.065_{-    0.014-    0.026}^{+    0.014+    0.027}$   \\

$n_s$ & $    0.9663_{-    0.0047-    0.0093}^{+    0.0047+    0.0091}$ & $    0.9677_{-    0.0038-    0.0075}^{+    0.0038+    0.0075}$  & $    0.9668_{-    0.0043-    0.0084}^{+    0.0043+    0.0084}$ & $    0.9722_{-    0.0043-    0.0086}^{+    0.0044+    0.0088}$ & $    0.9708_{-    0.0044-    0.0085}^{+    0.0043+    0.0088}$ & $    0.9666_{-    0.0045-    0.0087}^{+    0.0045+    0.0092}$  \\

${\rm{ln}}(10^{10} A_s)$ & $    3.093_{-    0.032-    0.066}^{+    0.033+    0.065}$ & $    3.097_{-    0.033-    0.062}^{+    0.033+    0.063}$  & $    3.097_{-    0.033-    0.065}^{+    0.033+    0.064}$  & $  3.113_{-    0.034-    0.065}^{+    0.033+    0.062}$  &  $    3.077_{-    0.032-    0.066}^{+    0.035+    0.065}$  & $    3.061_{-    0.025-    0.048}^{+    0.025+    0.049}$  \\

$\Omega_{m0}$ & $    0.312_{-    0.0097-    0.016}^{+    0.0085+    0.018}$ & $    0.309_{-    0.0062-    0.012}^{+    0.0061+    0.012}$  & $    0.311_{-    0.0085-    0.015}^{+    0.0078+    0.016}$ & $    0.298_{- 0.0075-    0.015}^{+    0.0075+    0.015}$ &  $    0.299_{-    0.0073-    0.014}^{+    0.0073+    0.014}$ & $    0.310_{-    0.0081-    0.016}^{+    0.0083+    0.017}$   \\

$\sigma_8$ & $    0.829_{-    0.013-    0.026}^{+    0.013+    0.026}$ & $    0.829_{-    0.013-    0.025}^{+    0.013+    0.026}$ & $    0.830_{-    0.013-    0.026}^{+    0.013+    0.026}$ & $    0.831_{-    0.013-    0.026}^{+    0.013+    0.025}$  &  $    0.816_{-    0.012-    0.025}^{+    0.012+    0.024}$  & $    0.815_{-    0.0087-    0.017}^{+    0.0086+    0.017}$ \\

$H_0$ & $   67.47_{-    0.65-    1.3}^{+    0.66+    1.2}$ & $   67.74_{-    0.46-    0.91}^{+    0.47+    0.89}$ & $   67.59_{-    0.60-    1.1}^{+    0.59+    1.2}$ & $   68.55_{-    0.58-    1.1}^{+ 0.59+    1.2}$ & $   68.50_{-    0.57-    1.1}^{+    0.57+    1.1}$ & $   67.64_{-    0.61-    1.2}^{+    0.60+    1.2}$  \\

$S_8$ & $0.846^{+ 0.018 + 0.037}_{- 0.017 - 0.034}$ &  
$0.841^{+ 0.015 + 0.031}_{- 0.015 -0.03}$ & $0.845^{+0.016 + 0.033}_{-0.017-0.033}$ & $0.828^{+0.014 + 0.03}_{-0.014 - 0.03}$ & $0.814^{+0.013 + 0.027}_{-0.013-0.026}$ & $0.828^{+ 0.012 + 0.025}_{-0.011 - 0.023}$\\

\hline 

$\chi^2$ & 12964.062 & 12969.178 & 13998.916 & 12980.808 & 13492.378 & 12973.924\\

\hline\hline                                                                                                                    
\end{tabular}
}                                                                                                                   
\caption{We report the 68\% and 95\% CL constraints on the free and derived parameters of the $\Lambda$CDM model using various observational datasets. In the last row of the table we also display the best-fit values of $\chi^2$ for all the observational datasets. }
\label{tab:LCDM}                                                                                                   
\end{table*}                                                                                                                     
\end{center}                                                                                                                    
\endgroup                                                           
\begin{figure*}
\includegraphics[width=0.9\textwidth]{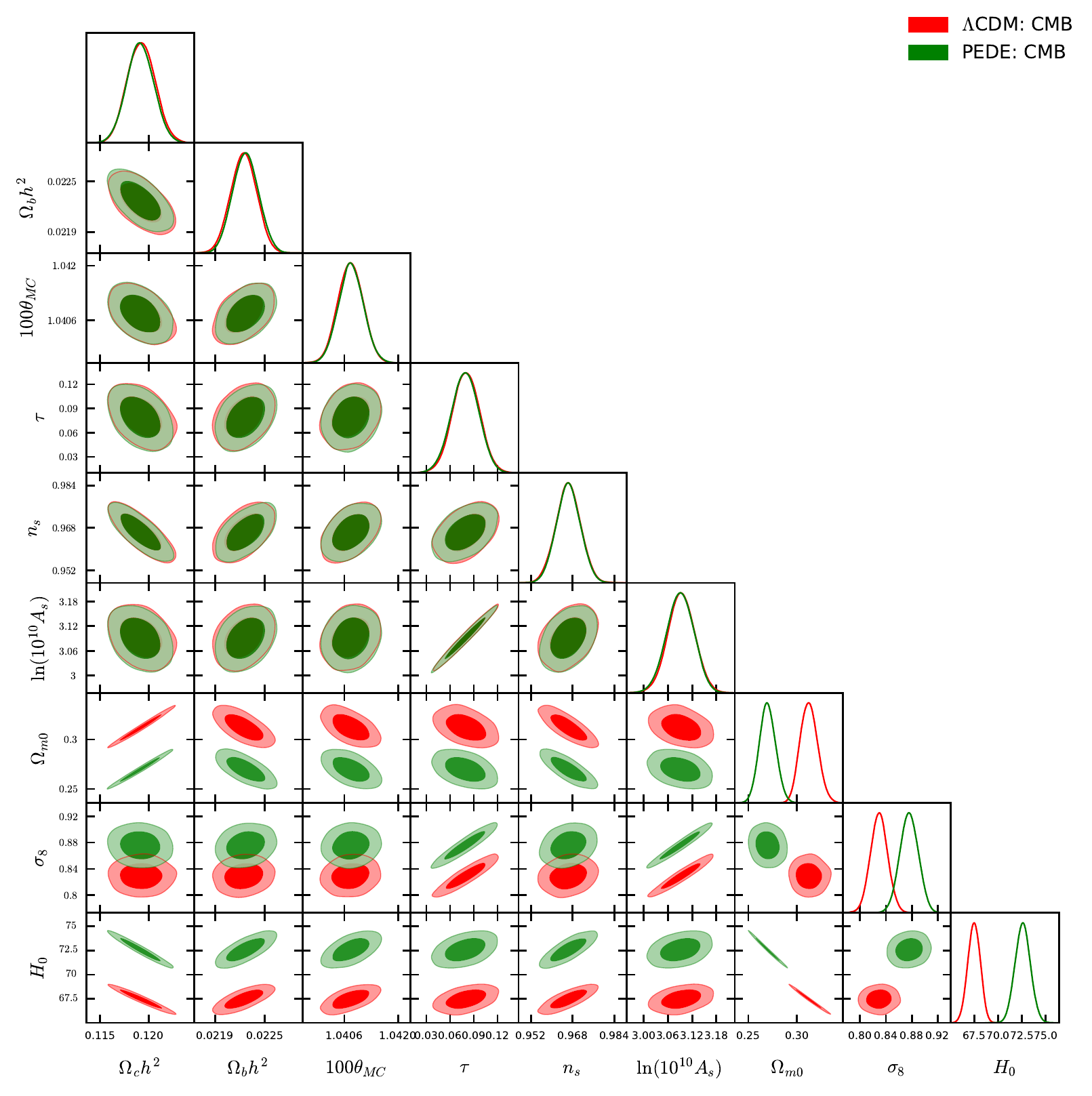}
\caption{We compare the observational constraints on PEDE and $\Lambda$CDM models obtained from the CMB data alone. }
\label{fig-cmb-comparison}
\end{figure*}
\begin{figure*}
\includegraphics[width=0.45\textwidth]{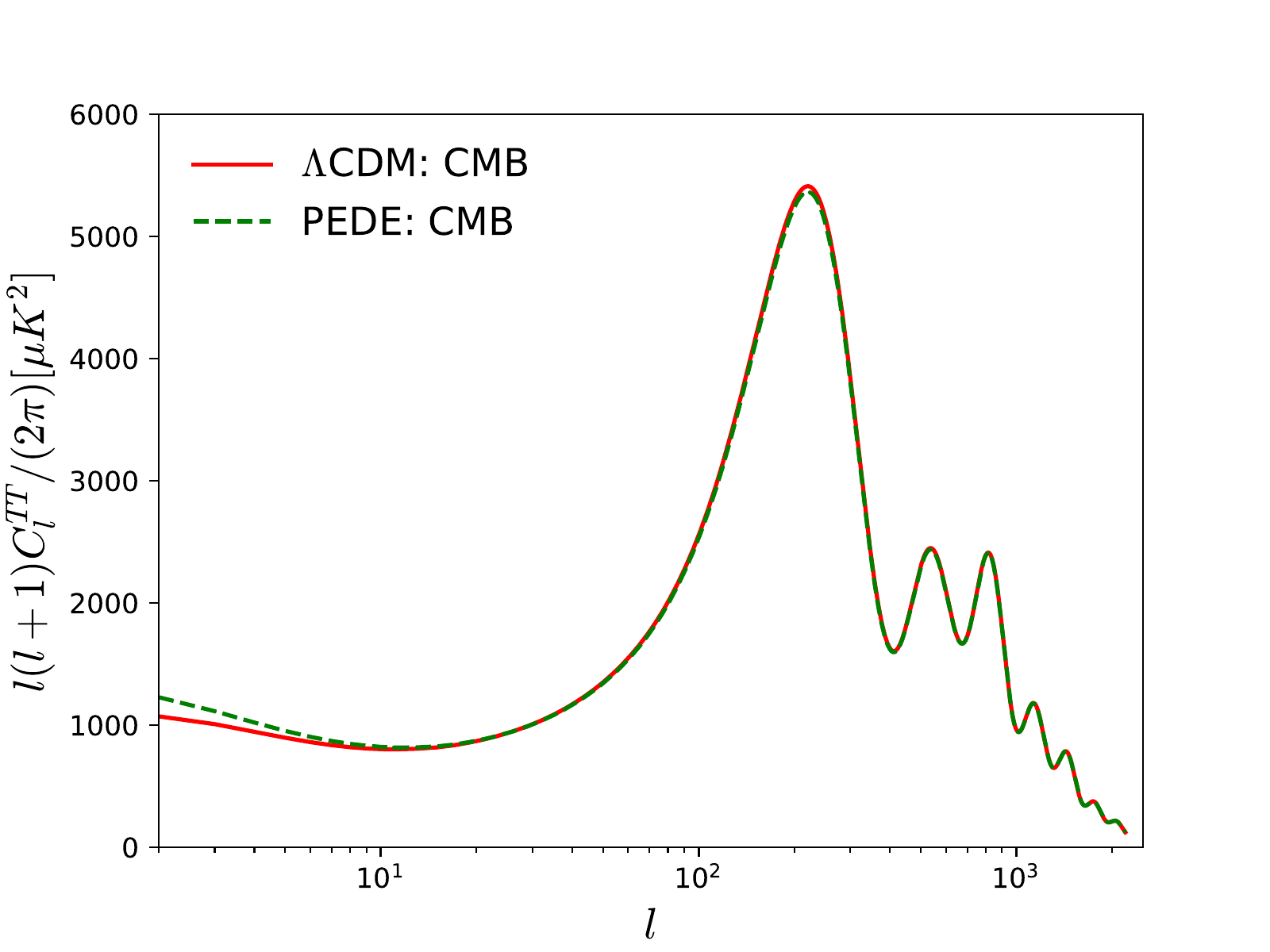}
\includegraphics[width=0.45\textwidth]{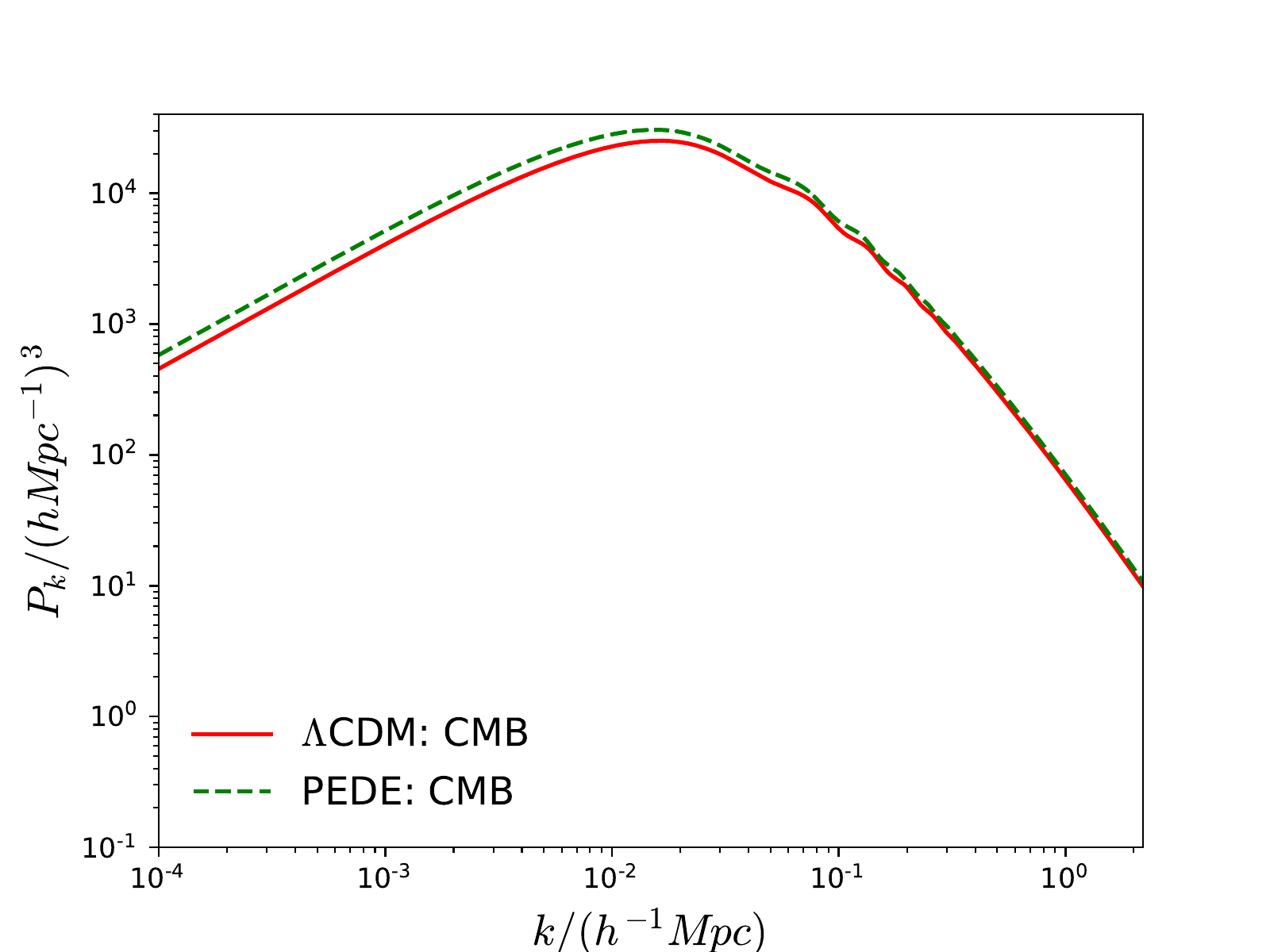}
\caption{We compare the CMB temperature power spectra (left graph) and the matter power spectra (right graph) computed for the PEDE and $\Lambda$CDM models taking the best-fit values of the model parameters summarized in Table \ref{tab:PEDE} and \ref{tab:LCDM}. }
\label{fig:cmb+matter}
\end{figure*}
                                                                                                        
  \begin{table*}
  \centering
  \begin{tabular}{|>{\bfseries}c|*{4}{c|}}\hline
    \multirow{2}{*}{Parameters} & \multicolumn{2}{c|}{ CMB+BAO+Pantheon} & \multicolumn{2}{c|}{ CMB+BAO+Pantheon+R19+DES+Lensing} \\\cline{2-5} & PEDE & $\Lambda$CDM & PEDE       & $\Lambda$CDM \\ \hline 
    
    $\Omega_c h^2$ & $    0.1220_{-    0.00094-    0.0018}^{+    0.00094+    0.0018}$ & $    0.1186_{-    0.0010-    0.0020}^{+    0.0010+    0.0019}$              & $    0.1201_{-    0.00082-    0.0016}^{+    0.00082+    0.0016}$              & $    0.1165_{-    0.00088-    0.0017}^{+    0.00087+    0.0017}$            \\ 
    
   $\Omega_b h^2$ & $    0.02206_{-    0.00013-    0.00026}^{+    0.00013+    0.00026}$ & $    0.02230_{-    0.00014-    0.00027}^{+    0.00014+    0.00027}$   & $    0.02216_{-    0.00013-    0.00025}^{+    0.00013+    0.00025}$             & $    0.02247_{-    0.00013-    0.00027}^{+    0.00013+    0.00026}$             \\ 
    
    $100\theta_{MC}$ & $    1.04044_{-    0.00031-    0.00062}^{+    0.00031+    0.00061}$ & $    1.04086_{-    0.00030-    0.00060}^{+    0.00031+    0.00059}$ & $    1.04063_{-    0.00029-    0.00057}^{+    0.00029+    0.00058}$  & $    1.04112_{-    0.00029-    0.00058}^{+    0.00030+    0.00059}$            \\ 
    
    $\tau$ & $    0.064_{-    0.016-    0.031}^{+    0.015+    0.031}$ & $    0.084_{-    0.016-    0.033}^{+    0.017+    0.032}$  & $    0.043_{-    0.011-    0.022}^{+    0.011+    0.022}$            & $    0.076_{-    0.012-    0.024}^{+    0.012+    0.025}$            \\ 
    
   $n_s$ & $    0.9589_{-    0.0037-    0.0073}^{+    0.0036+    0.0073}$ &  $    0.9679_{-    0.0039-    0.0074}^{+    0.0039+    0.0074}$ & $    0.9623_{-    0.0033-    0.0068}^{+    0.0034+    0.0068}$  & $    0.9723_{-    0.0037-    0.0072}^{+    0.0037+    0.0074}$ \\ 
   
    ${\rm{ln}}(10^{10} A_s)$ & $    3.068_{-    0.031-    0.062}^{+    0.031+    0.060}$ & $    3.100_{-    0.032-    0.066}^{+    0.033+    0.063}$ & $    3.019_{-    0.021-    0.043}^{+    0.021+    0.042}$  & $    3.078_{-    0.023-    0.045}^{+    0.023+    0.046}$           \\ 
    
    $\Omega_{m0}$ & $    0.287_{-    0.0058-    0.011}^{+    0.0057+    0.012}$ & $    0.308_{-    0.0060-    0.012}^{+    0.0060+    0.012}$   & $    0.276_{-    0.0049-    0.0094}^{+    0.0049+    0.0010}$ & $    0.295_{-    0.0051-    0.0095}^{+    0.0050+    0.010}$            \\ 
    
   $\sigma_8$ & $    0.873_{-    0.013-    0.027}^{+    0.013+    0.027}$ &  $    0.831_{-    0.013-    0.026}^{+    0.013+    0.027}$  & $    0.847_{-    0.0085-    0.017}^{+    0.0085+    0.017}$  & $    0.815_{-    0.0086-    0.017}^{+    0.0086+    0.017}$           \\ 
    
    $H_0$ & $   70.97_{-    0.52-    1.01}^{+    0.51+    0.99}$ & $   67.78_{-    0.45-    0.88}^{+    0.46+    0.91}$            & $   71.98_{-    0.46-    0.91}^{+    0.46+    0.89}$ & $   68.77_{-    0.40-    0.80}^{+    0.40+    0.79}$         \\ 
    
   $S_8$ & $0.854^{+0.015+0.031}_{-0.015-0.005}$ & $0.841^{+0.016+0.031}_{-0.015-0.03}$            & $    0.812_{-    0.0094-    0.018}^{+    0.0094+    0.018}$ & $    0.808_{-    0.0094-    0.018}^{+    0.0094+    0.018}$            \\ 
    
    \hline 
$\chi^2$ & 14020.478 & 14001.486 & 14563.008 & 14552.488  \\
\hline\hline                        
  \end{tabular}
  \caption{Summary of the observational constraints on the PEDE model and the $\Lambda$CDM model for the combined datasets CMB+BAO+Pantheon and CMB+BAO+Pantheon+R19+DES+Lensing.}
  \label{tab:CBP}   
  \end{table*}

\section{Results}
\label{sec-results}

The current PEDE model has the same number of free parameters as in spatially flat $\Lambda$CDM model. So, statistically within the spatially flat FLRW background, the PEDE and $\Lambda$CDM are on the same ground. We have constrained both the models using the same observational data (see section \ref{sec-data}) in order to perform a statistical comparison between them with the aim to focus on the tensions on both $H_0$ and $S_8 \equiv \sigma_8 \sqrt{\Omega_{m0}/3}$.

In Table \ref{tab:PEDE} we show the observational constraints on the PEDE model using a number of cosmological datasets such as CMB, CMB+BAO, CMB+Pantheon, CMB+R19, CMB+DES and CMB+Lensing. Fig.~\ref{fig:pede} shows the 1D posterior distributions of all parameters of this model together with 2D joint contours considering several combinations of the parameters at 68\% and 95\% CL. At the same time in order to make a comparison of the PEDE model with the $\Lambda$CDM cosmology, in Table \ref{tab:LCDM} we show the constraints on the $\Lambda$CDM scenario using the same combination of data of the PEDE model. Our analyses clearly show that for the PEDE model, the Hubble constant $H_0$, takes very high values compared to the values of $H_0$ obtained from $\Lambda$CDM model. For the PEDE model one can see that CMB dataset alone estimate, $H_0 = 72.58_{-0.80}^{+    0.79}$ km/s/Mpc (68\% CL) while for the same dataset $\Lambda$CDM model returns, $H_0 = 67.47_{- 0.65}^{+    0.66}$ km/s/Mpc (68\% CL). One can notice that the difference in the error bars on $H_0$ for both the models are not much significant, but the values of $H_0$ for 
the PEDE model is perfectly in agreement with the Hubble constant estimate from R19: $H_0=74.03\pm 1.42$ km/s/Mpc (68\% CL). Moreover, there is an improvement of the $\chi^2$ of about $1.5$ for the PEDE model with respect to the $\Lambda$CDM one for the same number of degrees of freedom. When external datasets, such as BAO, Pantheon, etc., are added to CMB dataset, the estimations of $H_0$ for all the observational combinations in the PEDE model (see Table \ref{tab:PEDE}), take significantly higher values compared to the $H_0$ estimations for $\Lambda$CDM one (see Table \ref{tab:LCDM}). Moreover, also the error bars on $H_0$ for PEDE model are really stable for all the observational datasets, therefore the $H_0$ tension reconciled within 68\% CL, for this PEDE model, is not due to a volume effect. {\it This is a very interesting result because without using any additional degrees of freedom, only dynamical character of the dark energy density (equivalently, the dark energy equation of state)  can reconcile the $H_0$ tension in a remarkable way.} One should note the symmetrical form of the dark energy density in this model that appears due to setting the pivot of transition which refers to the epoch of matter-dark energy density equality, to zero \footnote{However, the pivot of transition (matter-dark energy density
equality), $z_t$, can be considered as a derived parameter (depending on the value of matter density) by extending the present model (see the paragraph after eqn. (6) of \cite{Li:2019yem}) where the dark energy density parameter can be parametrized as,  $\Omega_{DE} = \Omega_{DE,0} \frac{F (z)}{F (z= 0)}$ in which $F (z) = 1- \tanh\left[\log_{10} (1+z) -  \log_{10} (1+z_t)\right]$. }.  Additionally, when R19 and DES are added to the CMB dataset, we see a large improvement of the $\chi^2$ for PEDE model compared to the $\Lambda$CDM model, for instance,  $\Delta \chi^2 \sim 17$ \footnote{Let us note that we define $\Delta \chi^2$ as: $\chi^2 ({\rm \Lambda CDM})- \chi^2 ({\rm PEDE})$} for CMB+R19 and $\Delta \chi^2 \sim 7$ for CMB+DES.  This large improvement we see is due to the fact that for these cases the CMB data are more in agreement with the additional data in the PEDE model with respect to the $\Lambda$CDM scenario. As one can see, for CMB alone case, $\chi^2$ for PEDE is less than the $\chi^2$ for $\Lambda$CDM model.  
For all the other combinations of data  (such as CMB+BAO, CMB+Pantheon and CMB+Lensing) the $\chi^2$ for PEDE gets worse compared to $\chi^2$ for $\Lambda$CDM. 

We present the comparisons between the CMB constraints of PEDE and $\Lambda$CDM model in Fig. \ref{fig-cmb-comparison}. We do not show other combinations because qualitatively they look similar. Here we can observe that all the cosmological parameters, with the exception of $H_0$, $\sigma_8$ and $\Omega_{m0}$, perfectly coincide in the PEDE and $\Lambda$CDM models. Instead, the Hubble constant $H_0$ and the clustering parameter $\sigma_8$ shift towards higher values, while $\Omega_{m0}$ towards a smaller one. 
If we now compute the $S_8$ parameter, the PEDE model seems to be able to alleviate also this tension, shifting $S_8$ more in agreement with the cosmic shear data. In fact, we found that in  PEDE model the DES alone estimates,  $S_8=0.848^{+0.023}_{-0.033}$ at 68$\%$ CL, in agreement within $1\sigma$ with the CMB. However, the tension between these two datasets in the PEDE model is not completely solved, because $\sigma_8$ is much higher for the CMB only than for DES only (for which $\sigma_8=0.665^{+0.030}_{-0.054}$ at 68$\%$ CL), and $\Omega_{m0}$ is much lower compared to its estimation from DES only:  $\Omega_{m0}=0.491_{-0.035}^{+0.045}$).

In Table~\ref{tab:CBP}, finally, we show the results for two different combinations, namely, the CMB+BAO+Pantheon combination and the full dataset CMB+BAO+Pantheon+R19+DES+Lensing, because, according to some recent works, see for example~\cite{Martinelli:2019krf}, it has been pointed out that the combination of all the three probes at the same time cannot alleviate the $H_0$ tension. We find that for CMB+BAO+Pantheon,  $H_0$ moves from $H_0=67.78^{+0.46}_{-0.45}$ km/s/Mpc at 68\% CL in the $\Lambda$CDM model to $H_0=70.97^{+0.51}_{-0.52}$ km/s/Mpc at 68\% CL in the PEDE model and for the CMB+BAO+Pantheon+R19+DES+Lensing combination, $H_0$ moves from $H_0=68.77_{- 0.40}^{+ 0.40}$ km/s/Mpc at 68\% CL in the $\Lambda$CDM model to $H_0=71.98_{- 0.46}^{+    0.46}$ km/s/Mpc at 68\% CL in the PEDE model. In other words, we can still alleviate the Hubble constant tension between CMB+BAO+Pantheon (CMB+BAO+Pantheon+R19+DES+Lensing) and R19, from 4.2 to 2 (from 3.6 to 1.4) standard deviations, compatible with a statistical fluke. However, this agreement happens by worsening the fit of the data, with a $\Delta \chi^2 \sim 19$ ($\Delta \chi^2 \sim 10.5 $). 

We now discuss the behaviour of this emergent DE model in the large scales through Fig.~\ref{fig:cmb+matter} where we explicitly compare the PEDE and $\Lambda$CDM models considering the CMB temperature anisotropy spectra and matter power spectra.  The left and right graph of Fig.~\ref{fig:cmb+matter}  respectively describe the CMB temperature anisotropy spectra and matter power spectra. 
From the left graph of Fig. \ref{fig:cmb+matter} we notice that at the lower multipoles (around $l\lesssim 10$), the PEDE has a slight deviation from the $\Lambda$CDM but such deviation is very mild and completely hidden by the cosmic variance. However, one can note that (see Fig. \ref{fig:cmb+matter}) the amplitude of the first acoustic peak in the CMB power-spectrum for both the models does not change at all.    
Similar observation can be found from the matter power spectra shown in the right side of Fig.~\ref{fig:cmb+matter}. So, PEDE has a mild deviation from the $\Lambda$CDM and this is only detected from the CMB and matter power spectra. 

\begin{table} [ht]             
\begin{tabular}{ccc}                
\hline\hline                                                        
$\ln B_{ij}$ &  Strength of evidence for model ${M}_i$ \\ \hline
$0 \leq \ln B_{ij} < 1$ & Weak \\
$1 \leq \ln B_{ij} < 3$ & Definite/Positive \\
$3 \leq \ln B_{ij} < 5$ & Strong \\
$\ln B_{ij} \geq 5$ & Very strong \\
\hline\hline                        
\end{tabular}                                                   \caption{The table shows the revised Jeffreys scale \cite{Kass:1995loi} which is used to compare the underlying cosmological models. } \label{tab:jeffreys}       
\end{table}      
\begin{table*} [!]
\begin{center}                    
\begin{tabular}{ccccccccc}                                      \hline\hline              
 
Dataset & $\ln B_{ij}$ & ~Strength of evidence\\

\hline
CMB  & $-0.2$ & Weak\\
CMB+BAO  & $-3.1$ & Strong\\
CMB+Pantheon  & $-5.8$ & Very Strong \\
CMB+R19  & $2.7$ & Definite/Positive\\
CMB+DES & $-1.6$ & Definite/Positive\\
CMB+Lensing & $-0.6$ & Weak\\
CMB+BAO+Pantheon & $-4.4$ &  Strong \\
CMB+BAO+Pantheon+R19+DES+Lensing & $-4.8$  & Strong \\
\hline\hline
\end{tabular}    
\caption{Summary of $\ln B_{ij}$ values, where $j$ refers to the reference model $\Lambda$CDM and $i$ refers to the PEDE model. The negative sign actually indicates that the reference model is favored over the PEDE model and the positive sign is for the reverse conclusion. } 
\label{tab:bayesian}                          
\end{center}    
\end{table*} 
Finally, we analyze the performance of the current PEDE model with respect to the standard $\Lambda$CDM model. It is a very natural question to ask how efficient a new cosmological model is, since from the theoretical ground, the introduction of a new dark energy model is very easy. So, we close this section with the 
Bayesian evidences computed for the PEDE model with respect to $\Lambda$CDM as the reference model. To calculate the Bayesian evidence for all the observational data we use a cosmological code \texttt{MCEvidence} originally developed by the authors of \cite{Heavens:2017afc,Heavens:2017hkr}. Let us note that the use of \texttt{MCEvidence} for computing the Bayesian evidences needs only the MCMC chains that are used to extract the cosmological parameters using the observational datasets (we also refer to~\cite{Pan:2017zoh,Yang:2018xah} for the same discussions). The performance of a cosmological model (say $M_i$) with respect to some reference cosmological model (here $\Lambda$CDM) is quantified through the Bayes factor $B_{ij}$ of the model $M_i$ with respect to the reference model $M_j$ (or, the logarithm of the Bayes factor, namely, $\ln B_{ij}$). In Table \ref{tab:jeffreys} we display the revised Jeffreys scale that quantifies the observational support of the underlying cosmological model and in Table~\ref{tab:bayesian} we summarize the values of $\ln B_{ij}$ computed for the PEDE model, for all the observational datasets. From the analysis, we clearly see that except from CMB+R19 combination, all other observational datasets favour $\Lambda$CDM over the PEDE. The interesting observation is the case with CMB+R19 where we see that PEDE is favored over $\Lambda$CDM with a positive evidence. This is in agreement with the observations because for CMB+R19, the $\chi^2$ for PEDE is much improved of about $~17$ compared to the $\chi^2$ for $\Lambda$CDM. This is also in agreement with the analyses of ~\cite{Li:2019yem} where the authors claim that the PEDE model can be favored compared to $\Lambda$CDM when some hard cut priors on $H_0$ is implemented. 

\section{Concluding remarks}
\label{sec-discuss}

Despite of having tremendous success to frame the presently ongoing accelerated expansion of the universe, the $\Lambda$-cosmology is equally challenged for several unexplained issues associated with it. The cosmological constant problem is undoubtedly one of the biggest challenges to explain. Apart from that the tensions in some parameters have been another remarkable issue at current time. The measurements of $H_0$ and $S_8$ in $\Lambda$CDM based framework do not agree with their measurements by other experimental missions $-$ known as tensions in the cosmological parameters.  The parameter $H_0$ is in more than $4\sigma$ tension between ($\Lambda$CDM-based) Planck and local observations by the SH0ES collaboration~\cite{Riess:2019cxk}. On the other hand, $S_8$ parameter is in tension between Planck and other observations, such as KiDS-450~\cite{Kuijken:2015vca,Hildebrandt:2016iqg,Conti:2016gav}, DES~\cite{Abbott:2017wau,Troxel:2017xyo} and CFHTLenS~\cite{Heymans:2012gg, Erben:2012zw,Joudaki:2016mvz}. 
Some recent literature investigating along this line found that an extended parameter space compared to $\Lambda$CDM is able to ease such tensions, however, due to extra free parameters, from Bayesian point of view, $\Lambda$CDM remains favored compared to the extended cosmological models. A natural inquiry, that forced us to look for an alternative cosmological model, having same number of parameters as in $\Lambda$CDM but with the potentiality to address some of the above problems.

A model proposed in ~\cite{Li:2019yem} seems to have such properties (having no degree of freedom for the dark energy sector) which influenced us to investigate this model further. In fact, in \cite{Li:2019yem} the authors presented the analyses at the level of background. Since the evolution of the model considering the large scale inhomogeneities is worth to provide a better picture of the model, hence, it is necessary to consider the perturbations. Therefore, keeping this important issue, we have investigated this model in a more comprehensive way and analysed how the model is able to reconcile the $H_0$ and $S_8$ tensions.

In Table~\ref{tab:PEDE} we show the observational constraints on the PEDE model using various cosmological datasets. In particular,  we have considered the analysis with CMB alone and the datasets in which one external dataset is included with CMB at a time. From Table~\ref{tab:PEDE} it is quite clear that $H_0$ takes considerably higher values compared to the estimations of $H_0$ for the $\Lambda$CDM model (see Table~\ref{tab:LCDM}). For CMB alone dataset, we see that  at 68\% CL, $H_0  = 72.58_{- 0.80}^{+    0.79}$ for the PEDE model which is pretty close to its local estimations by Riess et al. \cite{Riess:2019cxk} and the estimations for other datasets remain almost same with stable error bars on $H_0$. This clearly shows us that within 68\% CL, the tension on $H_0$ is perfectly reconciled, with an improvement of the $\chi^2$ for some certain combinations of the cosmological probes. {\it This is one of the very interesting findings because the PEDE model has exactly six parameters as in $\Lambda$CDM model.} In order to investigate further the ability of the model, we have constrained the PEDE (and also the $\Lambda$CDM model) using more cosmological probes at a time, such as CMB+BAO+Pantheon and CMB+BAO+Pantheon+R19+DES+Lensing, the results of which are summarized in Table~\ref{tab:CBP}. Our results clearly show that the tension on $H_0$ is still alleviated within 68\% CL, however, the alleviation in both the cases appear due to worsening the $\chi^2$ values compared to the $\Lambda$CDM model. 
Now concerning the  $S_8$ parameter,  we fond that its value using DES alone is in agreement with the estimations from CMB for PEDE model, so it is able to reconcile this tension as well. However, the tension between these two datasets (i.e. DES and CMB from Planck) in the PEDE model is not completely solved because we find that $\sigma_8$ is much higher for the CMB data alone compared to its estimation from DES alone, and additionally,  $\Omega_{m0}$ is much lower for CMB alone compared to its estimation from DES only.

In summary, it is evident that the current PEDE model is a new appealing addition in the literature of dark energy models which, based on its present observational features, should be considered as a potential candidate for further investigations. In a forthcoming work we plan to extend the present work by  including the non-gravitational interaction between dark matter and the dark energy having the equation of state explored in this work. The purpose to include the interaction within the present context is highly motivated because the interacting models have the ability to address some very important cosmological puzzles including the recently explored tensions in some cosmological parameters. We refer the readers to some of the works in this direction \cite{Yang:2018euj,Yang:2019vni,DiValentino:2019ffd,DiValentino:2019jae}. We aim to investigate the $H_0$ and $S_8$ tensions in order to see whether first of all the alleviation of  $H_0$ tension is independent of the interaction. And secondly, if the tension on $S_8$ is much relaxed compared to the present case study.

\section*{Acknowledgments}
The authors express their sincere thank to the referee for reading the work very carefully and for her/his comments and suggestions that certainly improved the quality of presentation of the article. SP has been supported by the Mathematical Research Impact-Centric Support Scheme (MATRICS), File No. MTR/2018/000940, given by the Science and Engineering Research Board (SERB), Govt. of India, as well as  by the Faculty Research and Professional Development Fund (FRPDF) Scheme of Presidency University, Kolkata, India.  WY acknowledges the support from the National Natural Science Foundation of China under Grants No.  
11705079 and No.  11647153. EDV acknowledges support from the European Research Council in the form of a Consolidator Grant with number 681431. AS would like to acknowledge the support of the Korea Institute for
Advanced Study (KIAS) grant funded by the Korea government.  SC acknowledges the Mathematical Research Impact Centric Support (MATRICS), project reference no. MTR/2017/000407, by the  Science and Engineering Research Board (SERB), Government of India.



\begin{thebibliography}{99}

\bibitem{Li:2019yem}   X.~Li and A.~Shafieloo,
  {\it Phenomenologically Emergent Dark Energy and Ruling Out Cosmological Constant,}
  Astrophys.\ J.\  {\bf 883}, no. 1, L3 (2019) 
  arXiv:1906.08275 [astro-ph.CO].

\bibitem{Ade:2015xua} 
  P.~A.~R.~Ade {\it et al.} [Planck Collaboration],
  {\it Planck 2015 results. XIII. Cosmological parameters,}
  Astron.\ Astrophys.\  {\bf 594}, A13 (2016)
  [arXiv:1502.01589 [astro-ph.CO]].
  
  \bibitem{Aghanim:2018eyx} 
  N.~Aghanim {\it et al.} [Planck Collaboration],
  {\it Planck 2018 results. VI. Cosmological parameters,}
  arXiv:1807.06209 [astro-ph.CO].
  
  \bibitem{Beutler:2011hx} 
  F.~Beutler {\it et al.},
  {\it The 6dF Galaxy Survey: Baryon Acoustic Oscillations and the Local Hubble Constant,}
  Mon.\ Not.\ Roy.\ Astron.\ Soc.\  {\bf 416}, 3017 (2011)
  [arXiv:1106.3366 [astro-ph.CO]].
  
  
\bibitem{Ross:2014qpa} 
  A.~J.~Ross, L.~Samushia, C.~Howlett, W.~J.~Percival, A.~Burden and M.~Manera,
  {\it The clustering of the SDSS DR7 main Galaxy sample - I. A 4 per cent distance measure at $z = 0.15$,}
  Mon.\ Not.\ Roy.\ Astron.\ Soc.\  {\bf 449}, no. 1, 835 (2015)
  [arXiv:1409.3242 [astro-ph.CO]].
  
  
\bibitem{Alam:2016hwk} 
  S.~Alam {\it et al.} [BOSS Collaboration],
  {\it The clustering of galaxies in the completed SDSS-III Baryon Oscillation Spectroscopic Survey: cosmological analysis of the DR12 galaxy sample,}
  Mon.\ Not.\ Roy.\ Astron.\ Soc.\  {\bf 470}, no. 3, 2617 (2017)
  [arXiv:1607.03155 [astro-ph.CO]].

\bibitem{Riess:2019cxk} 
  A.~G.~Riess, S.~Casertano, W.~Yuan, L.~M.~Macri and D.~Scolnic,
{\it Large Magellanic Cloud Cepheid Standards Provide a 1\% Foundation for the Determination of the Hubble Constant and Stronger Evidence for Physics beyond $\Lambda$CDM,}
  Astrophys.\ J.\  {\bf 876}, no. 1, 85 (2019)
  [arXiv:1903.07603 [astro-ph.CO]].  
  

\bibitem{Wong:2019kwg} 
  K.~C.~Wong {\it et al.},
  {\it H0LiCOW XIII. A 2.4\% measurement of $H_{0}$ from lensed quasars: $5.3\sigma$ tension between early and late-Universe probes,}
  arXiv:1907.04869 [astro-ph.CO].
  
 
\bibitem{Camarena:2019moy} 
  D.~Camarena and V.~Marra,
  {\it Cosmology-independent local determination of $H_0$ in strong tension with CMB,}
  arXiv:1906.11814 [astro-ph.CO].
  

\bibitem{Riess:2020sih} 
  A.~G.~Riess,
  {\it The Expansion of the Universe is Faster than Expected,}
  Nature Rev.\ Phys.\  {\bf 2}, no. 1, 10 (2019)
  [arXiv:2001.03624 [astro-ph.CO]].


\bibitem{Hildebrandt:2016iqg} 
  H.~Hildebrandt {\it et al.},
  {\it KiDS-450: Cosmological parameter constraints from tomographic weak gravitational lensing,}
  arXiv:1606.05338 [astro-ph.CO].
  
    
\bibitem{Kuijken:2015vca} 
  K.~Kuijken {\it et al.},
  {\it Gravitational Lensing Analysis of the Kilo Degree Survey,}
  Mon.\ Not.\ Roy.\ Astron.\ Soc.\  {\bf 454}, no. 4, 3500 (2015)
  [arXiv:1507.00738 [astro-ph.CO]].
  

\bibitem{Conti:2016gav} 
  I.~Fenech Conti, R.~Herbonnet, H.~Hoekstra, J.~Merten, L.~Miller and M.~Viola,
  {\it Calibration of weak-lensing shear in the Kilo-Degree Survey,}
  Mon.\ Not.\ Roy.\ Astron.\ Soc.\  {\bf 467}, no. 2, 1627 (2017)
  [arXiv:1606.05337 [astro-ph.CO]].
  
  \bibitem{Abbott:2017wau} 
  T.~M.~C.~Abbott {\it et al.} [DES Collaboration],
  {\it Dark Energy Survey Year 1 Results: Cosmological Constraints from Galaxy Clustering and Weak Lensing,}
  arXiv:1708.01530 [astro-ph.CO].
  
  
  \bibitem{Troxel:2017xyo} 
  M.~A.~Troxel {\it et al.} [DES Collaboration],
  {\it Dark Energy Survey Year 1 results: Cosmological constraints from cosmic shear,}
  Phys.\ Rev.\ D {\bf 98}, no. 4, 043528 (2018)
  [arXiv:1708.01538 [astro-ph.CO]].
  
  
 
  
\bibitem{Heymans:2012gg} 
  C.~Heymans {\it et al.},
  {\it CFHTLenS: The Canada-France-Hawaii Telescope Lensing Survey,}
  Mon.\ Not.\ Roy.\ Astron.\ Soc.\  {\bf 427}, 146 (2012)
  [arXiv:1210.0032 [astro-ph.CO]].


\bibitem{Erben:2012zw} 
  T.~Erben {\it et al.},
  {\it CFHTLenS: The Canada-France-Hawaii Telescope Lensing Survey - Imaging Data and Catalogue Products,}
  Mon.\ Not.\ Roy.\ Astron.\ Soc.\  {\bf 433}, 2545 (2013)
  [arXiv:1210.8156 [astro-ph.CO]].
  
  
\bibitem{Joudaki:2016mvz} 
  S.~Joudaki {\it et al.},
  {\it CFHTLenS revisited: assessing concordance with Planck including astrophysical systematics,}
  Mon.\ Not.\ Roy.\ Astron.\ Soc.\  {\bf 465}, no. 2, 2033 (2017)
  [arXiv:1601.05786 [astro-ph.CO]].
  
  
\bibitem{Palanque-Delabrouille:2019iyz} 
  N.~Palanque-Delabrouille, C.~Yèche, N.~Schöneberg, J.~Lesgourgues, M.~Walther, S.~Chabanier and E.~Armengaud,
  {\it Hints, neutrino bounds and WDM constraints from SDSS DR14 Lyman-$\alpha$ and Planck full-survey data,}
  arXiv:1911.09073 [astro-ph.CO].
  
  
 
\bibitem{Asgari:2019fkq} 
  M.~Asgari {\it et al.},
 {\it KiDS+VIKING-450 and DES-Y1 combined: Mitigating baryon feedback uncertainty with COSEBIs,}
  Astron.\ Astrophys.\  {\bf 634}, A127 (2020)
  [arXiv:1910.05336 [astro-ph.CO]].
  
  
\bibitem{Freedman:2020dne} 
  W.~L.~Freedman {\it et al.},
  arXiv:2002.01550 [astro-ph.GA].
  
  \bibitem{Hamana:2019etx} 
  T.~Hamana {\it et al.},
  {\it Cosmological constraints from cosmic shear two-point correlation functions with HSC survey first-year data,}
  arXiv:1906.06041 [astro-ph.CO].
  
  \bibitem{Mortsell:2018mfj} 
  E.~M\"{o}rtsell and S.~Dhawan,
  {\it Does the Hubble constant tension call for new physics?,}
  JCAP {\bf 1809}, no. 09, 025 (2018)
  [arXiv:1801.07260 [astro-ph.CO]].
  
  \bibitem{Vagnozzi:2019ezj} 
  S.~Vagnozzi,
  {\it New physics in light of the $H_0$ tension: an alternative view,}
  arXiv:1907.07569 [astro-ph.CO].
  
  
   
\bibitem{Efstathiou:2013via} 
  G.~Efstathiou,
  {\it H0 Revisited,}
  Mon.\ Not.\ Roy.\ Astron.\ Soc.\  {\bf 440}, no. 2, 1138 (2014)
  [arXiv:1311.3461 [astro-ph.CO]].
 


\bibitem{DiValentino:2015ola} 
  E.~Di Valentino, A.~Melchiorri and J.~Silk,
 {\it Beyond six parameters: extending $\Lambda$CDM,}
  Phys.\ Rev.\ D {\bf 92}, no. 12, 121302 (2015)
  [arXiv:1507.06646 [astro-ph.CO]].
  
\bibitem{DiValentino:2016hlg} 
  E.~Di Valentino, A.~Melchiorri and J.~Silk,
 {\it Reconciling Planck with the local value of $H_0$ in extended parameter space,}
  Phys.\ Lett.\ B {\bf 761}, 242 (2016)
  [arXiv:1606.00634 [astro-ph.CO]].
  
  \bibitem{Kumar:2017dnp} 
  S.~Kumar and R.~C.~Nunes,
  {\it Echo of interactions in the dark sector,} 
  Phys.\ Rev.\ D {\bf 96}, no. 10, 103511 (2017)
  [arXiv:1702.02143 [astro-ph.CO]].
  
  \bibitem{DiValentino:2017iww} 
  E.~Di Valentino, A.~Melchiorri and O.~Mena,
  {\it Can interacting dark energy solve the $H_0$ tension?,}
  Phys.\ Rev.\ D {\bf 96}, no. 4, 043503 (2017)
  [arXiv:1704.08342 [astro-ph.CO]].
  
  \bibitem{DiValentino:2017zyq} 
  E.~Di Valentino, A.~Melchiorri, E.~V.~Linder and J.~Silk,
 {\it Constraining Dark Energy Dynamics in Extended Parameter Space,}
  Phys.\ Rev.\ D {\bf 96}, no. 2, 023523 (2017)
  [arXiv:1704.00762 [astro-ph.CO]].
  
  \bibitem{Renk:2017rzu} 
  J.~Renk, M.~Zumalac\'{a}rregui, F.~Montanari and A.~Barreira,
  {\it Galileon gravity in light of ISW, CMB, BAO and H$_0$ data,}
  JCAP {\bf 1710}, no. 10, 020 (2017)
  [arXiv:1707.02263 [astro-ph.CO]].
  
 
\bibitem{Sola:2017znb} 
  J.~Sol\`{a},  A.~G\'{o}mez-Valent and J.~de Cruz P\'{e}rez,
  {\it The $H_0$ tension in light of vacuum dynamics in the Universe,}
  Phys.\ Lett.\ B {\bf 774}, 317 (2017)
  [arXiv:1705.06723 [astro-ph.CO]].
 
  
\bibitem{DiValentino:2017gzb} 
  E.~Di Valentino,
{\it Crack in the cosmological paradigm,}
  Nat.\ Astron.\  {\bf 1}, no. 9, 569 (2017)
  [arXiv:1709.04046 [physics.pop-ph]].
  
  \bibitem{DiValentino:2017oaw} 
  E.~Di Valentino, C.~B{\o}ehm, E.~Hivon and F.~R.~Bouchet,
  {\it Reducing the $H_0$ and $\sigma_8$ tensions with Dark Matter-neutrino interactions,}
  Phys.\ Rev.\ D {\bf 97}, no. 4, 043513 (2018)
  [arXiv:1710.02559 [astro-ph.CO]].
  

\bibitem{Fernandez-Arenas:2017isq} 
  D.~Fernandez Arenas {\it et al.},
  {\it An independent determination of the local Hubble constant},
  Mon.\ Not.\ Roy.\ Astron.\ Soc.\  {\bf 474}, no. 1, 1250 (2018)
  [arXiv:1710.05951 [astro-ph.CO]].
 
\bibitem{DiValentino:2017rcr} 
  E.~Di Valentino, E.~V.~Linder and A.~Melchiorri,
  {\it Vacuum phase transition solves the $H_0$ tension,}
  Phys.\ Rev.\ D {\bf 97}, no. 4, 043528 (2018)
  [arXiv:1710.02153 [astro-ph.CO]].
  
  \bibitem{Khosravi:2017hfi} 
  N.~Khosravi, S.~Baghram, N.~Afshordi and N.~Altamirano,
  {\it $H_0$ tension as a hint for a transition in gravitational theory,}
  Phys.\ Rev.\ D {\bf 99}, no. 10, 103526 (2019)
  [arXiv:1710.09366 [astro-ph.CO]].
  
 \bibitem{Nunes:2018xbm} 
  R.~C.~Nunes,
  {\it Structure formation in $f(T)$ gravity and a solution for $H_0$ tension,}
  JCAP {\bf 1805}, 052 (2018)
  [arXiv:1802.02281 [gr-qc]].
 
  \bibitem{Yang:2018euj} 
  W.~Yang, S.~Pan, E.~Di Valentino, R.~C.~Nunes, S.~Vagnozzi and D.~F.~Mota,
  {\it Tale of stable interacting dark energy, observational signatures, and the $H_0$  tension,}
  JCAP {\bf 1809}, no. 09, 019 (2018)
  [arXiv:1805.08252 [astro-ph.CO]].
  
  \bibitem{Colgain:2018wgk} 
  E.~\'{O} Colg\'{a}in, M.~H.~P.~M.~van Putten and H.~Yavartanoo,
  {\it de Sitter Swampland, $H_0$ tension \& observation,}
  Phys.\ Lett.\ B {\bf 793}, 126 (2019)
  [arXiv:1807.07451 [hep-th]].
  
  
  \bibitem{DEramo:2018vss} 
  F.~D'Eramo, R.~Z.~Ferreira, A.~Notari and J.~L.~Bernal,
  {\it Hot Axions and the $H_0$ tension,}
  JCAP {\bf 1811}, no. 11, 014 (2018)
  [arXiv:1808.07430 [hep-ph]].
  
  
\bibitem{Yang:2018uae} 
  W.~Yang, A.~Mukherjee, E.~Di Valentino and S.~Pan,
  {\it Interacting dark energy with time varying equation of state and the $H_0$ tension,}
  Phys.\ Rev.\ D {\bf 98}, no. 12, 123527 (2018)
  [arXiv:1809.06883 [astro-ph.CO]].
  
  \bibitem{Guo:2018ans} 
  R.~Y.~Guo, J.~F.~Zhang and X.~Zhang,
  {\it Can the $H_0$ tension be resolved in extensions to $\Lambda$CDM cosmology?,}
  JCAP {\bf 1902}, 054 (2019)
  [arXiv:1809.02340 [astro-ph.CO]].
  
  \bibitem{Yang:2018qmz}
  W.~Yang, S.~Pan, E.~Di Valentino, E.~N.~Saridakis and S.~Chakraborty, {\it Observational 
constraints on one-parameter dynamical dark-energy parametrizations and the $H_0$ tension,}  
Phys. Rev. D {\bf 99} no.4, 043543 (2019),
arXiv:1810.05141 [astro-ph.CO].
  
  \bibitem{Poulin:2018cxd} 
  V.~Poulin, T.~L.~Smith, T.~Karwal and M.~Kamionkowski,
 {\it Early Dark Energy Can Resolve The Hubble Tension,}
  Phys.\ Rev.\ Lett.\  {\bf 122}, no. 22, 221301 (2019)
  [arXiv:1811.04083 [astro-ph.CO]].
  
  \bibitem{Zhang:2018air} 
  X.~Zhang and Q.~G.~Huang,
  {\it Constraints on $H_0$ from WMAP and baryon acoustic osillation measurements,}
  arXiv:1812.01877 [astro-ph.CO].
  
 
\bibitem{Banihashemi:2018oxo} 
  A.~Banihashemi, N.~Khosravi and A.~H.~Shirazi,
  {\it Ups and Downs in Dark Energy: phase transition in dark sector as a proposal to lessen cosmological tensions,}
  arXiv:1808.02472 [astro-ph.CO].
  

\bibitem{Banihashemi:2018has} 
  A.~Banihashemi, N.~Khosravi and A.~H.~Shirazi,
  {\it Ginzburg-Landau Theory of Dark Energy: A Framework to Study Both Temporal and Spatial Cosmological Tensions Simultaneously,}
  Phys.\ Rev.\ D {\bf 99}, no. 8, 083509 (2019)
  [arXiv:1810.11007 [astro-ph.CO]].
  
  
 
 \bibitem{Kreisch:2019yzn} 
  C.~D.~Kreisch, F.~Y.~Cyr-Racine and O.~Dor\'{e},
  {\it The Neutrino Puzzle: Anomalies, Interactions, and Cosmological Tensions,}
  arXiv:1902.00534 [astro-ph.CO].
  
  \bibitem{Martinelli:2019dau} 
  M.~Martinelli, N.~B.~Hogg, S.~Peirone, M.~Bruni and D.~Wands,
 {\it Constraints on the interacting vacuum - geodesic CDM scenario,}
  arXiv:1902.10694 [astro-ph.CO].

  
  \bibitem{Vattis:2019efj} 
  K.~Vattis, S.~M.~Koushiappas and A.~Loeb,
 {\it Dark matter decaying in the late Universe can relieve the H0 tension,}
  Phys.\ Rev.\ D {\bf 99}, no. 12, 121302 (2019)
  [arXiv:1903.06220 [astro-ph.CO]].
  
  \bibitem{Kumar:2019wfs} 
  S.~Kumar, R.~C.~Nunes and S.~K.~Yadav,
  {\it Dark sector interaction: a remedy of the tensions between CMB and LSS data,}
  Eur.\ Phys.\ J.\ C {\bf 79}, no. 7, 576 (2019)
  [arXiv:1903.04865 [astro-ph.CO]].

\bibitem{Agrawal:2019lmo} 
  P.~Agrawal, F.~Y.~Cyr-Racine, D.~Pinner and L.~Randall,
  {\it Rock 'n' Roll Solutions to the Hubble Tension,}
  arXiv:1904.01016 [astro-ph.CO].
  
 \bibitem{Yang:2019jwn}
W.~Yang, S.~Pan, A.~Paliathanasis, S.~Ghosh and Y.~Wu,
{\it Observational constraints of a new unified dark fluid and the $H_0$ tension,}
Mon. Not. Roy. Astron. Soc. \textbf{490}, no.2, 2071-2085 (2019)
[arXiv:1904.10436 [gr-qc]].
  
 \bibitem{Yang:2019qza}
W.~Yang, S.~Pan, E.~Di Valentino, A.~Paliathanasis and J.~Lu,
{\it Challenging bulk viscous unified scenarios with cosmological observations,}
Phys. Rev. D \textbf{100}, no.10, 103518 (2019)
[arXiv:1906.04162 [astro-ph.CO]].
 
\bibitem{Yang:2019uzo}
W.~Yang, O.~Mena, S.~Pan and E.~Di Valentino,
{\it Dark sectors with dynamical coupling,}
Phys. Rev. D \textbf{100}, no.8, 083509 (2019)
[arXiv:1906.11697 [astro-ph.CO]].

\bibitem{DiValentino:2019exe}
E.~Di Valentino, R.~Z.~Ferreira, L.~Visinelli and U.~Danielsson,
{\it Late time transitions in the quintessence field and the $H_0$ tension,}
Phys. Dark Univ. \textbf{26}, 100385 (2019)
[arXiv:1906.11255 [astro-ph.CO]].
  
  
  \bibitem{Desmond:2019ygn} 
  H.~Desmond, B.~Jain and J.~Sakstein,
  {\it A local resolution of the Hubble tension: The impact of screened fifth forces on the cosmic distance ladder,}
  arXiv:1907.03778 [astro-ph.CO].
  
  \bibitem{Yang:2019nhz}
W.~Yang, S.~Pan, S.~Vagnozzi, E.~Di Valentino, D.~F.~Mota and S.~Capozziello,
{\it Dawn of the dark: unified dark sectors and the EDGES Cosmic Dawn 21-cm signal,}
JCAP \textbf{11}, 044 (2019)
[arXiv:1907.05344 [astro-ph.CO]].
  
 
 \bibitem{Pan:2019gop}
S.~Pan, W.~Yang, E.~Di Valentino, E.~N.~Saridakis and S.~Chakraborty,
{\it Interacting scenarios with dynamical dark energy: Observational constraints and alleviation of the $H_0$ tension,}
Phys. Rev. D \textbf{100}, no.10, 103520 (2019)
[arXiv:1907.07540 [astro-ph.CO]].
  
  \bibitem{Visinelli:2019qqu} 
  L.~Visinelli, S.~Vagnozzi and U.~Danielsson,
  {\it Revisiting a negative cosmological constant in light of low-redshift data,}
  arXiv:1907.07953 [astro-ph.CO].
  
  
\bibitem{Martinelli:2019krf} 
  M.~Martinelli and I.~Tutusaus,
  {\it CMB tensions with low-redshift $H_0$ and $S_8$ measurements: impact of a redshift-dependent type-Ia supernovae intrinsic luminosity,}
  arXiv:1906.09189 [astro-ph.CO].
  
  
  \bibitem{Cai:2019bdh} 
  Y.~F.~Cai, M.~Khurshudyan and E.~N.~Saridakis,
  {\it Model-independent reconstruction of $f(T)$ gravity from Gaussian Processes and alleviation of the $H_0$ tension,}
  arXiv:1907.10813 [astro-ph.CO].
  
  \bibitem{Schoneberg:2019wmt} 
  N.~Sch\"{o}neberg, J.~Lesgourgues and D.~C.~Hooper,
  {\it The BAO+BBN take on the Hubble tension,}
  arXiv:1907.11594 [astro-ph.CO].
  

\bibitem{Shafieloo:2016bpk} 
  A.~Shafieloo, D.~K.~Hazra, V.~Sahni and A.~A.~Starobinsky,
{\it Metastable Dark Energy with Radioactive-like Decay,}
  Mon.\ Not.\ Roy.\ Astron.\ Soc.\  {\bf 473}, no. 2, 2760 (2018)
  [arXiv:1610.05192 [astro-ph.CO]].
  

\bibitem{Li:2019san} 
  X.~L.~Li, A.~Shafieloo, V.~Sahni and A.~A.~Starobinsky,
  {\it Revisiting Metastable Dark Energy and Tensions in the Estimation of Cosmological Parameters,}
  arXiv:1904.03790 [astro-ph.CO].
  

  
  \bibitem{Pourtsidou:2016ico} 
  A.~Pourtsidou and T.~Tram,
  {\it Reconciling CMB and structure growth measurements with dark energy interactions,}
  Phys.\ Rev.\ D {\bf 94}, no. 4, 043518 (2016)
  [arXiv:1604.04222 [astro-ph.CO]].

\bibitem{An:2017crg} 
  R.~An, C.~Feng and B.~Wang,
  {\it Relieving the Tension between Weak Lensing and Cosmic Microwave Background with 
Interacting Dark Matter and Dark Energy Models,}
  JCAP {\bf 1802}, no. 02, 038 (2018)
  [arXiv:1711.06799 [astro-ph.CO]].
  
  
\bibitem{Gomez-Valent:2017idt} 
  A.~G\'{o}mez-Valent and J.~Sol\`{a},
  {\it Relaxing the $\sigma_8$-tension through running vacuum in the Universe,}
  EPL {\bf 120}, no. 3, 39001 (2017)
  [arXiv:1711.00692 [astro-ph.CO]].
  
  
  \bibitem{DiValentino:2018gcu} 
  E.~Di Valentino and S.~Bridle,
  {\it Exploring the Tension between Current Cosmic Microwave Background and Cosmic Shear 
Data,} Symmetry {\bf 10}, no. 11, 585 (2018).
  
  
  
\bibitem{Gomez-Valent:2018nib} 
  A.~G\'{o}mez-Valent and J.~Sol\`{a} Peracaula,
  {\it Density perturbations for running vacuum: a successful approach to structure formation and to the $\sigma_8$-tension,}
  Mon.\ Not.\ Roy.\ Astron.\ Soc.\  {\bf 478}, no. 1, 126 (2018)
  [arXiv:1801.08501 [astro-ph.CO]].
  
  \bibitem{Kazantzidis:2018rnb} 
  L.~Kazantzidis and L.~Perivolaropoulos,
  {\it Evolution of the $f\sigma_8$ tension with the Planck15/$\Lambda$CDM determination and implications for modified gravity theories,}
  Phys.\ Rev.\ D {\bf 97}, no. 10, 103503 (2018)
  [arXiv:1803.01337 [astro-ph.CO]].
   
  
\bibitem{Hazra:2018opk} 
  D.~K.~Hazra, A.~Shafieloo and T.~Souradeep,
  {\it Parameter discordance in Planck CMB and low-redshift measurements: projection in the primordial power spectrum,}
  JCAP {\bf 1904}, no. 04, 036 (2019)
  [arXiv:1810.08101 [astro-ph.CO]].
  
  \bibitem{Kazantzidis:2019dvk} 
  L.~Kazantzidis and L.~Perivolaropoulos,
  {\it Is gravity getting weaker at low z? Observational evidence and theoretical implications,}
  arXiv:1907.03176 [astro-ph.CO].
  
  \bibitem{Macaulay:2013swa} 
  E.~Macaulay, I.~K.~Wehus and H.~K.~Eriksen,
  {\it Lower Growth Rate from Recent Redshift Space Distortion Measurements than Expected from Planck,}
  Phys.\ Rev.\ Lett.\  {\bf 111}, no. 16, 161301 (2013)
  [arXiv:1303.6583 [astro-ph.CO]].
  
  
  \bibitem{Ma:1995ey} 
  C.~P.~Ma and E.~Bertschinger,
  {\it Cosmological perturbation theory in the synchronous and conformal Newtonian gauges,}
  Astrophys.\ J.\  {\bf 455}, 7 (1995)
  [astro-ph/9506072].
  
  \bibitem{Arendse:2019hev} 
  N.~Arendse {\it et al.},
  {\it Cosmic dissonance: new physics or systematics behind a short sound horizon?,}
  arXiv:1909.07986 [astro-ph.CO].


\bibitem{Palanque-Delabrouille:2015pga}
  N.~Palanque-Delabrouille {\it et al.},
 {\it Neutrino masses and cosmology with Lyman-alpha forest power spectrum,}
  JCAP {\bf 1511}, no.11,  011 (2015) 
  [arXiv:1506.05976 [astro-ph.CO]].
  
\bibitem{Giusarma:2016phn}
  E.~Giusarma, M.~Gerbino, O.~Mena, S.~Vagnozzi, S.~Ho and K.~Freese,
  {\it Improvement of cosmological neutrino mass bounds,}
  Phys.\ Rev.\ D {\bf 94}, no.8,  083522 (2016) 
  [arXiv:1605.04320 [astro-ph.CO]].
  
\bibitem{Vagnozzi:2017ovm}
  S.~Vagnozzi, E.~Giusarma, O.~Mena, K.~Freese, M.~Gerbino, S.~Ho and M.~Lattanzi,
  {\it Unveiling $\nu$ secrets with cosmological data: neutrino masses and mass hierarchy,}
  Phys.\ Rev.\ D {\bf 96} no.12,  123503 (2017) 
  [arXiv:1701.08172 [astro-ph.CO]].
  
\bibitem{Giusarma:2018jei}
  E.~Giusarma, S.~Vagnozzi, S.~Ho, S.~Ferraro, K.~Freese, R.~Kamen-Rubio and K.~B.~Luk,
  {\it Scale-dependent galaxy bias, CMB lensing-galaxy cross-correlation, and neutrino masses,}
  Phys.\ Rev.\ D {\bf 98}  no.12,  123526 (2018)
  [arXiv:1802.08694 [astro-ph.CO]].
 
 
 
  
\bibitem{Adam:2015rua} 
  R.~Adam {\it et al.} [Planck Collaboration],
  {\it Planck 2015 results. I. Overview of products and scientific results,}
  Astron.\ Astrophys.\  {\bf 594}, A1 (2016)
  [arXiv:1502.01582 [astro-ph.CO]].


\bibitem{Aghanim:2015xee} 
  N.~Aghanim {\it et al.} [Planck Collaboration],
  {\it Planck 2015 results. XI. CMB power spectra, likelihoods, and robustness of parameters,}
  Astron.\ Astrophys.\  {\bf 594}, A11 (2016)
  [arXiv:1507.02704 [astro-ph.CO]].

  

  
  

  
  \bibitem{Scolnic:2017caz} 
  D.~M.~Scolnic {\it et al.},
  {\it The Complete Light-curve Sample of Spectroscopically Confirmed SNe Ia from Pan-STARRS1 and Cosmological Constraints from the Combined Pantheon Sample,}
  Astrophys.\ J.\  {\bf 859}, no. 2, 101 (2018)
  [arXiv:1710.00845 [astro-ph.CO]].
  
  

  
\bibitem{Krause:2017ekm} 
  E.~Krause {\it et al.} [DES Collaboration],
  {\it Dark Energy Survey Year 1 Results: Multi-Probe Methodology and Simulated Likelihood Analyses,}
  [arXiv:1706.09359 [astro-ph.CO]].
  
  

\bibitem{Ade:2015zua} 
  P.~A.~R.~Ade {\it et al.} [Planck Collaboration],
  {\it Planck 2015 results. XV. Gravitational lensing,}
  Astron.\ Astrophys.\  {\bf 594}, A15 (2016)
  [arXiv:1502.01591 [astro-ph.CO]].
  
   
\bibitem{Lewis:2002ah} 
  A.~Lewis and S.~Bridle,
  {\it Cosmological parameters from CMB and other data: A Monte Carlo approach,}
  Phys.\ Rev.\ D {\bf 66}, 103511 (2002)
  [astro-ph/0205436].


\bibitem{Lewis:1999bs} 
  A.~Lewis, A.~Challinor and A.~Lasenby,
  {\it Efficient computation of CMB anisotropies in closed FRW models,}
  Astrophys.\ J.\  {\bf 538}, 473 (2000)
  [astro-ph/9911177].
  
  \bibitem{Heavens:2017afc} 
  A.~Heavens, Y.~Fantaye, A.~Mootoovaloo, H.~Eggers, Z.~Hosenie, S.~Kroon and 
E.~Sellentin,
  {\it Marginal Likelihoods from Monte Carlo Markov Chains},
  arXiv:1704.03472 [stat.CO].
  
  
  \bibitem{Heavens:2017hkr} 
  A.~Heavens, Y.~Fantaye, E.~Sellentin, H.~Eggers, Z.~Hosenie, S.~Kroon and 
A.~Mootoovaloo,
  {\it No evidence for extensions to the standard cosmological model},
  Phys.\ Rev.\ Lett.\  {\bf 119}, no. 10, 101301 (2017) 
[arXiv:1704.03467 [astro-ph.CO]].


\bibitem{Pan:2017zoh} 
  S.~Pan, E.~N.~Saridakis and W.~Yang,
  {\it Observational Constraints on Oscillating Dark-Energy Parametrizations,}
  Phys.\ Rev.\ D {\bf 98}, no. 6, 063510 (2018)
  [arXiv:1712.05746 [astro-ph.CO]].
  
  \bibitem{Yang:2018xah} 
  W.~Yang, M.~Shahalam, B.~Pal, S.~Pan and A.~Wang,
  {\it Constraints on quintessence scalar field models using cosmological observations,}
  Phys.\ Rev.\ D {\bf 100}, no. 2, 023522 (2019)
  [arXiv:1810.08586 [gr-qc]].
  
  \bibitem{Kass:1995loi}
  R.~E.~Kass and A.~E.~Raftery,
  {\it Bayes Factors,}
  J.\ Am.\ Statist.\ Assoc.\  {\bf 90}, no.430,  773 (1995). 
  
  
 \bibitem{Yang:2019vni} 
  W.~Yang, S.~Vagnozzi, E.~Di Valentino, R.~C.~Nunes, S.~Pan and D.~F.~Mota,
  {\it Listening to the sound of dark sector interactions with gravitational wave standard sirens,}
  JCAP {\bf 1907}, 037 (2019)
  [arXiv:1905.08286 [astro-ph.CO]]. 
 
 \bibitem{DiValentino:2019ffd} 
  E.~Di Valentino, A.~Melchiorri, O.~Mena and S.~Vagnozzi,
  {\it Interacting dark energy after the latest Planck, DES, and $H_0$ measurements: an excellent solution to the $H_0$ and cosmic shear tensions,}
  arXiv:1908.04281 [astro-ph.CO].

\bibitem{DiValentino:2019jae}
E.~Di Valentino, A.~Melchiorri, O.~Mena and S.~Vagnozzi,
{\it Nonminimal dark sector physics and cosmological tensions,}
Phys. Rev. D \textbf{101}, no.6, 063502 (2020)
[arXiv:1910.09853 [astro-ph.CO]].
    
\end{thebibliography}
\end{document}